\documentclass[letterpaper,11pt]{article}
\usepackage{geometry} 
\usepackage[utf8]{inputenc}
\usepackage[T1]{fontenc}
\usepackage{lmodern}

\usepackage[dvipsnames,table]{xcolor}
\usepackage{graphicx}
\graphicspath{{./}{figures/}}
\usepackage{amsmath}
\usepackage{hyperref} 
\hypersetup{colorlinks = true,
            linkcolor=black,
            urlcolor=RoyalBlue,
            citecolor=black,
            }
\usepackage[nameinlink,noabbrev]{cleveref}
\usepackage{url}
\usepackage{gensymb}
\usepackage{fancyhdr}
\usepackage{wrapfig}
\usepackage{sidecap}
\usepackage{natbib}  
\usepackage{amssymb}
\usepackage{tcolorbox}
\usepackage{longtable}
\usepackage{enumitem}
\usepackage{floatrow}
\newfloatcommand{capbtabbox}{table}[][\FBwidth]
\usepackage{lipsum}
\usepackage[displaymath, mathlines]{lineno}
\modulolinenumbers[2]
\usepackage{float}
\usepackage{pdfpages} 
\usepackage{todonotes}
\usepackage{sectsty}
\usepackage{authblk}

\sectionfont{\fontsize{12}{15}\selectfont}
\subsectionfont{\fontsize{12}{15}\selectfont}

\definecolor{lightgray}{gray}{0.9}
\date{\vspace{-5ex}}

\begin{document}

\noindent
\textbf{\large Roman CCS White Paper}

\vspace{2em}
\begin{center}
{\Large Options to Increase the Coverage Area of Prism Time Series in the High-Latitude Time Domain Core Community Survey}
\end{center}

\vspace{1.5em}
\noindent
\textbf{Roman Core Community Survey:} High Latitude Time Domain Survey

\vspace{0.5em}
\noindent
\textbf{Scientific Categories:} stellar physics and stellar types; stellar populations and the interstellar medium; large scale structure of the universe

\vspace{0.5em}
\noindent
\textbf{Additional scientific keywords:} Supernovae, Population III stars, Cosmology, Dark energy

\vspace{1em}
\noindent
\textbf{Submitting Author:}\\
Benjamin Rose, Baylor University (Ben\_Rose@baylor.edu)

\vspace{0.5em}
\noindent
\textbf{List of contributing authors:}\\
Sebastian Gomez, STScI (sgomez@stsci.edu)\\
Rebekah Hounsell, University of Maryland Baltimore County, NASA Goddard Space Flight Center (rebekah.a.hounsell@nasa.gov)\\
Bhavin Joshi, Johns Hopkins University (bjoshi5@jhu.edu)\\
David Rubin, Univserity of Hawaii (drubin@hawaii.edu)\\
Dan Scolnic, Duke University (dan.scolnic@duke.edu)\\
Masao Sako, University of Pennsylvania (masao@sas.upenn.edu)

\vspace{1em}
\noindent
\textbf{Abstract:} \\
The current reference High-latitude time domain survey increases the completeness of transients with prism temporal time series data by adjusting the ratio of prism-to-imaging time. However, there are two other nobs that allow for a more complete prism coverage: prism cadence and exposure time. In this white paper, we discuss how changes to the prism cadence and exposure time---in order to increase the fraction of observed transients with spectral time series---affect supernova cosmology, transient typing and template building, and the study of rare transients.

\thispagestyle{empty}
\newpage
\setcounter{page}{1}


\section{Introduction}

The current design reference survey for the High-latitude Time Domain Survey includes a yet to be determined amount of time dedicated to slitless spectroscopy with the P127 prism \citep{Rose2021c}. The prism data will be used for several scientific analyses including,
Type Ia supernova (SN~Ia) standardization, transient typing, template building, and redshift measurements to name a few.
In order to get a larger fraction of transients covered with spectral time-series data, the current reference survey increases the time dedicated to spectroscopy.
In this white paper, we investigate how to increase the percentage of transients with spectral time-series while keeping the same ratio of prism to imaging time. 


Assuming a fixed time of prism observations (say 1.5 months as used in the first reference mission presented in \citealt{Rose2021c}), the survey area can be increased by either reducing the frequency of visits or reducing the per-visit exposure time. The suitability of each is dependent on the noise profile and how your science is impacted by temporal information. 
Changing either the exposure time or cadence of these would affect all scientific analyses with the prism data, but transient science cases are the main drivers.
For this reason, we will look at how changes to the survey cadence effect spectroscopy of transient sources, not host galaxy redshifts.

\section{Exposure Time}

In the design reference missions, the prism observations will be Poisson limited. This is true even at 1/5 of the current exposure time, or the same total survey exposure time but with a 1-day cadence. 
As commuted in a simple exposure time calculator, the total survey exposure time is not dramatically increased by reducing the prism cadence while keeping the same total exposure time per five days. The same 5-day cadence depths can be obtained by stacking the 1-day observations. Though, some temporal information and co-added signal-to-noise would be lost for faint objects, this observation strategy would increase the temporal resolution of bright objects, compared to the design reference missions.


\section{Cadence}

By using separate cadences for photometry and spectroscopy, we would be able to reduce the spectroscopy cadence in a way that adds additional spectral field sizes. In \citet{Rose2021c}, only about 20\% of the photometric pointings have spectroscopy (for the survey that uses 25\% of the survey for spectroscopy). If we reduce the cadence by a factor of 2, the areas could be increased so that 40\% of the survey has both photometry and spectroscopy without increasing the overall fraction of spectroscopic observation time.
However, this would come at a cost to temporal information.

\section{Scientific Impacts}

\textbf{Dark Energy Figure of Merit}~The current Dark Energy figure of merit (FoM) estimates (like the ones in \citealt{Hounsell2018}, \citealt{Rose2021c}) assume a photometric centric SN Ia analysis---such as the ones done in Pantheon+ \citep{Brout2022b} and DES \citep[][Vincenzi et al., in prep.]{Brout2019}. These do not directly use the SN Ia spectral data \citep{Fakhouri2015} or use a spectral standardization model \citep{Saunders2018,Boone2021a,Stein2022}. Though, it is expected that these could improve both the statistical and systematic uncertainties on SN Ia distances \citep{Boone2021b}. Therefore, the prism data could improve the Dark Energy FoM. 




\vspace{1em}\noindent
\textbf{Transient Typing}~Transient typing and characterization is typically done with single high signal-to-noise spectroscopy. Therefore, spectra at max brightness is preferred. Though there are good classifiers that work on photometric data \citep[e.g.,][]{Boone2019,Moller2020, Qu2021,Boone2021c}, they require training and therefore enough labeled examples (from spectroscopic classification) of rare transients.

\vspace{1em}\noindent
\textbf{SN Ia Templates}~SN Ia standardization using light curves requires the creation of a generative model. Most resent cosmological models use some variant of the SALT model \citep{Guy2007,Kenworthy2021}. Currently, this model is trained by a program called \texttt{SALTshaker} \citep{Kenworthy2021}. For this process, spectra are used to facilitate K-correction-like extrapolations, allowing for the use of the large photometric data sets where the rest frame spectral energy distribution is sampled by the photometry slightly differently by each SN~Ia. Therefore, a subsample of spectral time-series is required to build a spectral model from the main photometric data set.

\vspace{1em}\noindent
\textbf{CCSN Templates}~Core-collapse SN (CCSN) templates are useful for a wide range of science cases \citep[e.g.,][]{Vincenzi2019}. Specifically, this field needs early-time behavior, since this is dominated by explosion physics and is difficult to extrapolate from later phases. A sufficiently fast cadence and sufficiently large data set, will ensure that Roman spectroscopy can drastically improve our CCSN templates.

\vspace{1em}\noindent
\textbf{Rare Transients}~Another key use of spectral time series is to understand the astrophysics of rare transients. These can be Kilonovae or even Population III stellar explosions (see CCS white papers by Ori Fox and Sebastian Gomez). This is a wide range of transients that evolve on observer timescales of a few days to several months. Assuming Roman's prims cadence is sufficiently fast, i.e. not longer than the timescale of the fastest transient one wants to observe, rare transients benefit from the largest area possible. Currently, the Roman High-latitude Time Domain Survey is only expected to observe a few Kilonoave over 20 square degrees (\citealt{Scolnic2018}, and Rose et al., in prep.). Therefore, anything that can increase the area of the prism observations, the more likely Roman will get a spectrum of a Kilonovae without using a target of opportunity.

\section{Conclusion}
The prism will provide information to a wide range of science cases. Most of these cases can be improved by a larger area of prism observations. This can be achieved by adjusting the cadence and exposure times, in addition to the prism-imaging time ratio, as was explored in previous High-latitude Time Domain reference surveys \citep{Rose2021c}.

\bibliographystyle{apj}
\bibliography{library}

\end{document}